\documentclass[fleqn,usenatbib]{mnras}

\usepackage[T1]{fontenc}
\usepackage{ae,aecompl}

\DeclareRobustCommand{\VAN}[3]{#2}
\let\VANthebibliography\thebibliography
\def\thebibliography{\DeclareRobustCommand{\VAN}[3]{##3}\VANthebibliography}

\usepackage{graphicx}	
\usepackage{amsmath}	
\usepackage{amssymb}	

\usepackage{multicol}
\usepackage[normalem]{ulem}
\usepackage{bm}
\usepackage{amsfonts}
\usepackage{graphicx}
\usepackage{color}
\usepackage{orcidlink}

\definecolor{grey}{rgb}{0.75,0.75,0.75}
\definecolor{Orange}{rgb}{1.0,0.5,0.15}
\definecolor{brown}{rgb}{0.7,0.25,0.0}
\definecolor{pink}{rgb}{1.0,0.5,0.5}
\definecolor{darkerred}{rgb}{0.8,0,0}
\definecolor{darkerblue}{rgb}{0,0,0.8}
\definecolor{Blue}{rgb}{0,0.08,0.65}
\definecolor{Red}{rgb}{0.65,0.08,0.05}
\definecolor{Green}{rgb}{0.15,0.45,0.25}

\newcommand{\pd}{\partial}
\newcommand{\rcm}{\bm{r}_{\mathrm{cm}}}
\newcommand{\qcm}{\bm{q}_{\mathrm{cm}}}
\newcommand{\cm}{\mathrm{cm}}

\newcommand{\dd}{\mathrm{d}}

\newcommand{\kk}{\bm{k}}
\newcommand{\rr}{\bm{r}}
\newcommand{\qq}{\bm{q}}
\newcommand{\xx}{\bm{x}}

\newcommand{\vv}{\bm{v}}

\newcommand{\tr}{\mathrm{tr}}

\title[Protohalo shear]{The energy shear of protohaloes}

\author[M. Musso, G. Despali and R. K. Sheth]{
Marcello Musso \orcidlink{0000-0003-1640-2795}$^{1}$\thanks{E-mail: mmusso@usal.es},
Giulia Despali \orcidlink{0000-0001-6150-4112}$^{2,3,4}$,
and Ravi K. Sheth \orcidlink{0000-0002-2330-0917}$^{5,6}$\thanks{E-mail: shethrk@upenn.edu}
\\
$^{1}$Departamento de F\'{\i}sica Fundamental and IUFFyM,
Universidad de Salamanca, E-37008 Salamanca, Spain\\
$^{2}$ Dipartimento di Fisica e Astronomia, Alma Mater Studiorum Università di Bologna, via Gobetti 93/2, I-40129 Bologna, Italy \\
$^{3}$ INAF-Osservatorio di Astrofisica e Scienza dello Spazio di Bologna, Via Piero Gobetti 93/3, I-40129 Bologna, Italy\\
$^{4}$  INFN-Sezione di Bologna, Viale Berti Pichat 6/2, I-40127 Bologna, Italy\\
$^{5}$Center for Particle Cosmology, University of Pennsylvania, 209 S. 33rd St., Philadelphia, PA 19104, USA\\
$^{6}$The Abdus Salam International Center for Theoretical Physics, Strada Costiera, 11, Trieste 34151, Italy
}

\date{Accepted XXX. Received YYY; in original form ZZZ}

\pubyear{2024}

\begin{document}
\label{firstpage}
\pagerange{\pageref{firstpage}--\pageref{lastpage}}
\maketitle

\begin{abstract}
As it collapses to form a halo, the shape of a protohalo patch is deformed by the initial shear field.  This deformation is often modeled using the `deformation' tensor, constructed from second derivatives of the gravitational potential, whose trace gives the initial overdensity.  However, especially for lower mass protohalos, this matrix is not always positive definite:  One of its eigenvalues has a different sign from the others.  We show that the evolution of a patch is better described by the `energy shear' tensor, which {\em is} positive definite and plays a direct role in the evolution.  This positive-definiteness simplifies models of halo abundances, assembly and of the cosmic web.  
\end{abstract}




\section{Introduction}
Understanding how dark matter halos form and cluster -- how their abundance and spatial distribution evolves -- potentially unlocks a wealth of information about cosmology and provides the scaffolding on which models of galaxy formation are built.  In addition, the intrinsic distribution of halo shapes is a leading systematic for constraining cosmological models with the next generation of galaxy cluster surveys, so halo formation has been the subject of significant work.  

Numerical simulations have shown that the protohalo patches, which evolve into virialized halos at later times, form from special regions in the initial fluctuation field. Following \cite{bbks86}, most models of halo formation associate protohalo patches with peaks in the smoothed overdensity field, which then `collapse' due to gravitational instability.  The simplest models assumed the patch and its collapse were both spherically symmetric \citep{gg72}, with later work exploring ellipsoidal rather than spherical collapse \citep{bm96, smt01}.  

These models typically distinguish between the `shape' and the `deformation' or `tidal' tensor of the initial patch. The latter uses the anisotropy of the second spatial derivatives of the potential to infer how the initial shape is deformed.  However, to approximate the initial shape, it is usual to use the tensor of second spatial derivatives of the density field. Since the two tensors are correlated, this proxy for the protohalo shape predicts preferential alignment of the shortest axis (corresponding to the steepest descent direction) with the direction of fastest infall \citep{vdwb96}: i.e., in these models, an initial sphere becomes increasingly non-spherical if the deformation tensor is anistropic.  However, if one uses the actual positions of the particles that make up the patch to define the initial inertia tensor, then one finds that the initial shapes are not spherical:  the longest axis of the inertia tensor is very well-aligned with the direction of fastest infall \citep{PorcianiTTTal2002,dts13, porciani11}.  
This alignment has a transparent physical interpretation: it allows particles from further away to reach the halo center at about the same time as particles that were initially less distant, but were collapsing towards the center more slowly \cite[also see][]{eshape23}.
The residual misalignment of the two tensors is instead what gives rise to torques/angular momentum \citep{whiteTTT1984,PorcianiTTTspin2002,CadiouSmith2021}.

The word `collapse' suggests that the object shrinks, in comoving coordinates, as it evolves.  Stated more carefully, the principal axes of the inertia tensor are expected to shrink as they are squeezed by the velocity flows that define the deformation tensor.  Squeezing suggests that the principal axes of the deformation tensor all have the same sign.  Indeed, some have used the requirement that all three axes have the same sign as the guiding principle from which to build a model of halo abundances \citep{ls98}.  However, simulations have shown that this is simply not true for a significant fraction of lower mass protohalo patches \citep{dts13}.  

Some problems also arise when working with peaks in the matter overdensity field -- the trace of the deformation tensor -- to identify protohalo centers. There is no guarantee that the density peak coincides with the position onto which the local gravitational flow converges, which is a natural choice to identify the geometric center of the protohalo patch.  Moreover, some of the integrals required to self-consistently define peak statistics diverge.  Both these disadvantages are overcome if one works instead with peaks in the energy overdensity field \citep{epeaks}, rather then in the matter overdensity.  The main goal of the present work is to show that the energy analogue of the deformation tensor, which we refer to as the `energy shear' tensor, also has an appealing property: all its eigenvalues have the same sign. 

We motivate why this sign requirement is expected in Section~\ref{sec:thy}, and show tests in simulations in Section~\ref{sec:sims}.  A final section discusses our findings and conclusions.  


\section{The energy overdensity tensor}\label{sec:thy}

In this Section, we study the conditions required for the triaxial collapse of protohalo patches of arbitrary shape, before discussing how they relate to analytical models typically based on spherically averaged quantities.

\subsection{Evolution of the inertia tensor of protohaloes}\label{sec:evolveI}

The formation of a dark matter halo can be described as the collapse of all three axes of its inertia tensor
\begin{equation}
  I_{ij}\equiv\int_V\dd^3r\rho(\rr,t)(r_i-r_{\cm,i})(r_j-r_{\cm,j})\,,
\label{eq:Iij_def}
\end{equation}
where $V$ is a freely evolving volume (in the physical coordinates $\rr$) containing the conserved mass $M$, and $\rcm$ denotes its center of mass. Mass conservation guarantees that time derivatives ``go through'' the integration sign and the time dependence of $\rho(\rr,t)$. Thus, the evolution of the inertia tensor of a region comoving with the fluid (the so-called virial equation) reads
\begin{equation}
  \frac{\ddot I_{ij}}{2} = 2K_{ij} + \frac{1}{2}(U_{ij}+U_{ji})
  + \frac{\Lambda}{3} I_{ij}
\label{eq:virial}
\end{equation}
\cite[see e.g.][for the $\Lambda=0$ case]{chandra,BinneyTremaineBook1987}, where $K_{ij}$ and $U_{ij}$ are the kinetic and potential energy tensors of the body. They are defined as
\begin{align}
  K_{ij} &\equiv \frac{1}{2}\int_V\dd^3r\rho(\rr,t) (\dot r_i-\dot r_{\cm,i})(\dot r_j-\dot r_{\cm,j})\,,\\
  U_{ij} &\equiv-\int_V\dd^3r\rho(\rr,t) (r_i- r_{\cm,i})(\nabla_j\Phi-[\nabla_j\Phi]_{\cm,j})\,,
\end{align}
where $\nabla\Phi$ is the gravitational attraction due to matter, so that 
 $\ddot\rr = -\nabla\Phi + (\Lambda/3)\rr$, 
and $[\nabla\Phi]_\cm$ is the one of the center of mass.

To describe the evolution of $I_{ij}$ with respect to the background, we split the matter gravitational potential as $\nabla\Phi = 4\pi G\bar\rho(\rr/3+\nabla\phi)$, where $\phi$ is the potential perturbation obeying $\nabla^2\phi = \delta_m$, and $\delta_m = (\rho/\bar\rho)-1$ is the matter density perturbation. We then rewrite equation \eqref{eq:virial} as
\begin{equation}
  \bigg(\frac{I_{ij}}{a^2}\bigg)^{\!\cdot\cdot}   
  + 2H \bigg(\frac{I_{ij}}{a^2}\bigg)^{\!\cdot}
  = 4k_{ij} - \frac{4\pi G\bar\rho}{3}\,(u_{ij}+u_{ji})
  \frac{I}{a^2}\,,
\label{eq:pertvir}
\end{equation}
where $a$ is the scale factor, $H=\dot a/a$ is the Hubble parameter, $I \equiv I_{kk}$ is the trace of $I_{ij}$,
\begin{equation}
  k_{ij} \equiv \frac{1}{2}\int_V\dd^3r\rho(\rr,t)
  \bigg(\frac{r_i- r_{\cm,i}}{a}\bigg)^{\!\cdot}
  \bigg(\frac{r_j-r_{\cm,j}}{a}\bigg)^{\!\cdot}
\end{equation}
is the peculiar kinetic energy tensor, and
\begin{equation}
  u_{ij} \equiv \frac{3}{I}\int_V\dd^3r\rho(\rr,t) (r_i- r_{\cm,i})(\nabla_j\phi-[\nabla_j\phi]_{\cm,j})\,,
  \label{eq:uij_def}
\end{equation}
is the potential energy overdensity tensor, or the `energy shear'. 
The cosmological constant does not appear explicitly in equation \eqref{eq:pertvir}; it only affects the perturbations through its effect on the background evolution of $\bar\rho$ and $H$.

Parameterizing the initial value of the comoving inertia tensor as $I_{ij}/a^2=(M/5)(AA^T)_{ij}$, equation \eqref{eq:pertvir} is solved by
\begin{equation}
  \frac{I_{ij}}{a^2}\simeq \frac{M}{5}\bigg[ (AA^T)_{ij}
  -D(z) (u_{ij}^\mathrm{L}+u_{ji}^\mathrm{L})\frac{\tr(AA^T)}{3}\bigg]
  \label{eq:Iij}
\end{equation}
at first order in $D(z)$, the growth function of matter density perturbations,
since $k_{ij}$ is automatically of second order. In this expression, $u_{ij}^\mathrm{L}$ is the linearized version of $u_{ij}$ from equation \eqref{eq:uij_def} divided by $D$, and is therefore time independent. For ease of notation, in the following  we will drop the superscript L and simply assume that all quantities are evaluated at the lowest order in density perturbations, and rescaled by $D(z)$.

This result can also be obtained directly by writing equation \eqref{eq:Iij_def} in Lagrangian coordinates as
\begin{equation}
    I_{ij} = a^5\bar\rho \int_V\dd^3q\, (q_i+\Psi_i)\,(q_j+\Psi_j)\,,
\end{equation}
where $\rr -\rcm = a(\qq + \mathbf{\Psi}(\qq,t))$ and $\mathbf{\Psi}$ is the displacement relative to the center of mass from the initial comoving coordinate $\qq$. At first order in Lagrangian perturbation theory, since $\mathbf{\Psi} \simeq -D(\mathbf{\nabla}\phi(\qq)-[\nabla\phi]_\cm)$, one gets equation \eqref{eq:Iij}.

While non-linear dynamics may modify the subsequent evolution, linear theory is sufficient to predict if and when each axis decouples from the Hubble flow and starts recollapsing. Since non-linear corrections, even large ones, manifest themselves at small scales, they will not be able to reverse an ongoing collapse. Hence, the condition for triaxial collapse to happen is that (the symmetric part of) $u_{ij}$ be positive definite:  i.e., all its eigenvalues have the same sign.

Taking the trace of equation \eqref{eq:Iij} returns
\begin{equation}
  R_I^2\equiv\frac{5I}{3M} \simeq a^2\frac{\tr(AA^T)}{3}\left[1-2D(z)\frac{\epsilon}{3}\right]\,,
\end{equation}
with $\epsilon\equiv\tr(u)$, which corresponds (at first order in perturbations) to the result that the collapse of the inertial radius $R_I$ is well described by spherical collapse with overdensity $\epsilon$. This is the basis for identifying protohalos
with local maxima of the energy overdensity \citep{epeaks}, since these are minima of the collapse time of $R_I$.

Naively, for a halo to form, one might imagine that all the three axes of the inertia tensor $I_{ij}$ should collapse at the same time as $R_I$. Equation \eqref{eq:Iij} would then require that
\begin{equation}
    AA^T\simeq \frac{u+u^T}{2\epsilon}\tr(AA^T)\,,
\end{equation}
in which case one would have
\begin{equation}
    I_{ij} \simeq \frac{M}{5}a^2\frac{u+u^T}{2\epsilon}\tr(AA^T)
    \left(1-2D\frac{\epsilon}{3}\right) 
    \simeq \frac{u+u^T}{2\epsilon} I\,,
    \label{eq:align}
\end{equation}
predicting that the eigenvectors of $I_{ij}$ and of $u_{ij}+u_{ji}$ are to be aligned, and their eigenvalues proportional. 
This picture is not entirely correct: we know in fact that protohalo boundaries tend to follow equipotential surfaces \citep{eshape23}, which gives an independent prescription for $I_{ij}$. It must however be true to some degree, at least for the eigenvectors, since equation \eqref{eq:Iij} does show that the anisotropy of $u_{ij}$ favours infall from the direction of strongest compression, with slower infall in the direction of weakest compression.

On the other hand, the mean deformation tensor (sometimes also called shear, or tidal tensor when referring to its traceless part) is defined as
\begin{equation}
    q_{ij} \equiv \frac{1}{M}\int_V\dd^3r\,\rho(\rr,t)\,
    \nabla_i\nabla_j\phi\,.
    \label{eq:deften}
\end{equation}
At leading order in perturbations, when $\rho(\rr,t)\simeq\bar\rho$, $q_{ij}$ is equal to the gradient of the center of mass acceleration $\nabla_i[\nabla_j\phi]_\cm$, and its trace, the volume average of $\nabla^2\phi$, equals the mean matter overdensity. 
Note that although the trace of $q_{ij}$ determines the evolution of the volume $V$, the tensor $q_{ij}$ does {\em not} otherwise play a direct role in the evolution of $I_{ij}$:  what matters is $u_{ij}$, rather than $q_{ij}$ (cf. equation~\ref{eq:Iij}).

\subsection{Spherical initial volumes}\label{sec:spheres}

When building models, one often approximates protohaloes as spheres of the same mass. If $V$ is a sphere of radius $R$ centered at $\xx$, containing mass $M=(4\pi/3)\bar\rho R^3$, at leading order in density perturbations its energy overdensity $\epsilon_R$ is
\begin{equation}
    \epsilon_R(\xx) = \int\frac{\dd^3k}{(2\pi)^3}\,\delta_\mathrm{L}(\kk)\, W_2(kR)\, e^{i\kk\cdot\xx}
    \label{eq:epsR}
\end{equation}
where $\delta_\mathrm{L}(k)$ is the linear matter overdensity field normalized to $\sigma_8$, and $W_2(kR) \equiv 15j_2(kR)/(kR)^2$ \citep{epeaks}. 
Similarly, its energy shear is given by
\begin{equation}
    u_{R,ij}(\xx) = \int\frac{\dd^3k}{(2\pi)^3}\,\delta_\mathrm{L}(\kk)\, 
    \frac{k_ik_j}{k^2}\, W_2(kR) \, e^{i\kk\cdot\xx}\,,
    \label{eq:uRij}
\end{equation}
whose trace is equation~(\ref{eq:epsR}).   
Peaks in the energy overdensity field are locations where $\nabla_i\epsilon_R=0$ and the Hessian
\begin{equation}
    \nabla_i\nabla_j\epsilon_R =
    - \int\frac{\dd^3k}{(2\pi)^3}\, \delta_\mathrm{L}(\kk)\, k_ik_j\, W_2(kR) \, e^{i\kk\cdot\xx}
    \label{eq:Hesseps}
\end{equation}
is negative definite.

For a sphere, the mean mass overdensity $\delta_R$, its deformation tensor $q_{R,ij}$ (whose trace is $\delta_R$) and Hessian $\nabla_i\nabla_j\delta_R$ are obtained replacing $W_2(kR)$ in equations \eqref{eq:epsR}, \eqref{eq:Hesseps} and \eqref{eq:uRij} with the standard Top Hat filter $W_1(kR)\equiv 3j_1(kR)/kR$.  Hence, while the difference between quantities associated with energy rather than mass density is conceptually rather important, in Fourier space the difference simply boils down to using a different, more UV-suppressed smoothing of the overdensity field $\delta(\kk)$ than the usual Top-Hat one.

The distributions of these Gaussian variates are characterized by the spectral moments 
\begin{equation}
    \sigma_{jn}^2\equiv\int\frac{\dd k}{k}\,\frac{k^3P(k)}{2\pi^2}\,k^{2j}\,W_n^2(kR)\,,
\label{eq:moments}
\end{equation}
where, following \cite{epeaks}, 
\begin{equation}
    W_n(x)\equiv(2n+1)!! \, \frac{j_n(x)}{x^n}\,.
\end{equation}
In particular, $\sigma_{02}$ and $\sigma_{22}$ are the standard deviations of $\epsilon_R$ and $\nabla^2\epsilon_R$, whereas $\sigma_{01}$ and $\sigma_{21}$ are those of $\delta_R$ and $\nabla^2\delta_R$.

For a homogeneous spherical density perturbation, and only in this case, the two quantities are equal: $\epsilon_R = \delta_R$, and the traceless parts of $u_{ij}$ and $q_{ij}$ are both zero. This is because only the $k=0$ mode of $\delta(\kk)$ matters in this case, but both filters tend to 1 as $k\to0$, so they make no difference.

\subsection{Quantifying the amount of anisotropy}
\label{sec:qu}

The anisotropic strength of the protohalo environment can be quantified by the magnitude of the traceless shear 
\begin{align}
  q_u^2 &\equiv \frac{3}{2}\tr(\bar u\bar u^T) 
      = \frac{3}{2}(u_{ij} - \epsilon\delta_{ij}/3)(u_{ji} - \epsilon\delta_{ji}/3)\nonumber\\
     &= \frac{(\lambda_1-\lambda_2)^2}{2} + \frac{(\lambda_1-\lambda_3)^2}{2} 
      + \frac{(\lambda_2-\lambda_3)^2}{2}\,.
  \label{eq:qu2}
\end{align}
(The suffix $u$ is to distinguish it from the analogous variable that can be defined from the traceless part of the deformation tensor, which is often called $q$ because it contributes a quadrupolar signature in perturbation theory.)

For a random Gaussian matrix, and therefore for random, unconstrained positions in a Gaussian field, this random variate would not be correlated with $\epsilon$ (the trace of $u_{ij}$) and would follow a $\chi^2$-distribution with five degrees of freedom \citep{st02}.
However, it is easy to see that simply requesting the energy shear to be positive definite induces a correlation between the two variables.  In fact, since
\begin{align}
  &\epsilon^2-q_u^2 = 3(\lambda_1\lambda_2 + \lambda_1\lambda_3 + \lambda_2\lambda_2) 
\end{align}
and the right hand side is positive when all eigenvalues are positive, if $u_{ij}$ is positive definite one has that $\epsilon>q_u$. Furthermore, the limit can be saturated only if all three eigenvalues are zero, which never happens in practice.

Some intuitive understanding of this correlation can be obtained by noticing that if $\epsilon$ is small, then all eigenvalues are likely to be equally small (since they all have the same sign and can no longer cancel out from their sum). Hence, their differences must be small as well.  Below, we look for this correlation in the statistics of protohaloes.

\section{Simulation measurements}\label{sec:sims}

\begin{figure*}
  \includegraphics[width=\columnwidth]{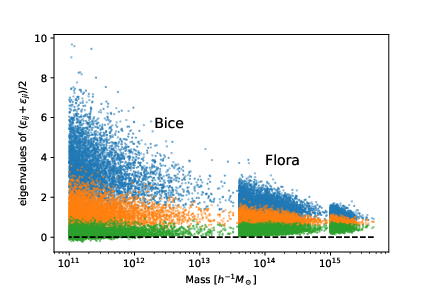}
  \includegraphics[width=\columnwidth]{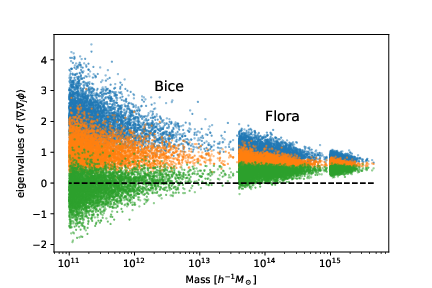}
\caption{\emph{Left panel.} Eigenvalues of the potential energy overdensity tensor $u_{ij}$. Only 5\% of protohaloes have one negative eigenvalue; this is our main result. 
\emph{Right panel.} Eigenvalues of the mean deformation tensor. The lowest eigenvalue is negative in about 40\% of the all protohaloes, and in more than 50\% of the lower mass ones, in stark contrast to the left panel.} 
\label{fig:eigenvalues}
  \includegraphics[width=\columnwidth]{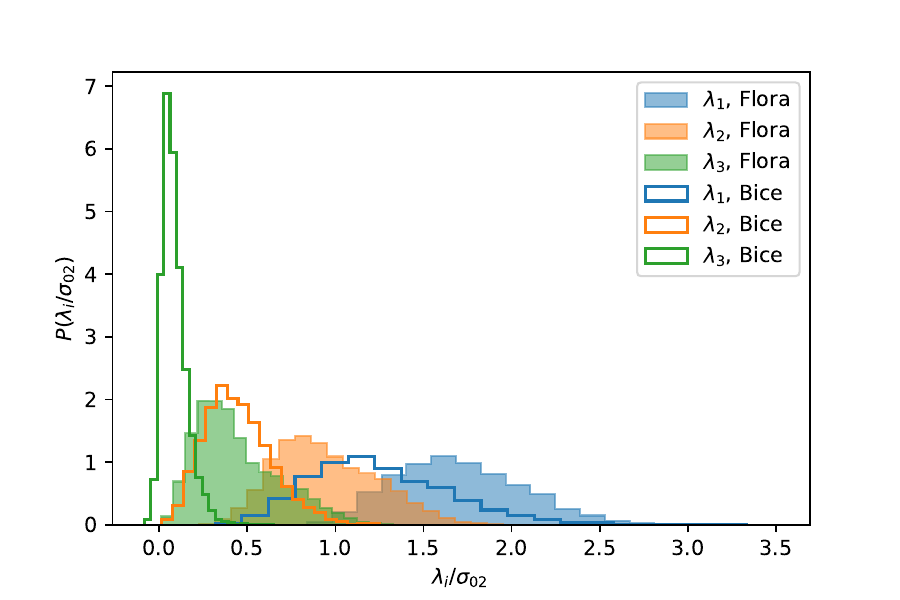}
  \includegraphics[width=\columnwidth]{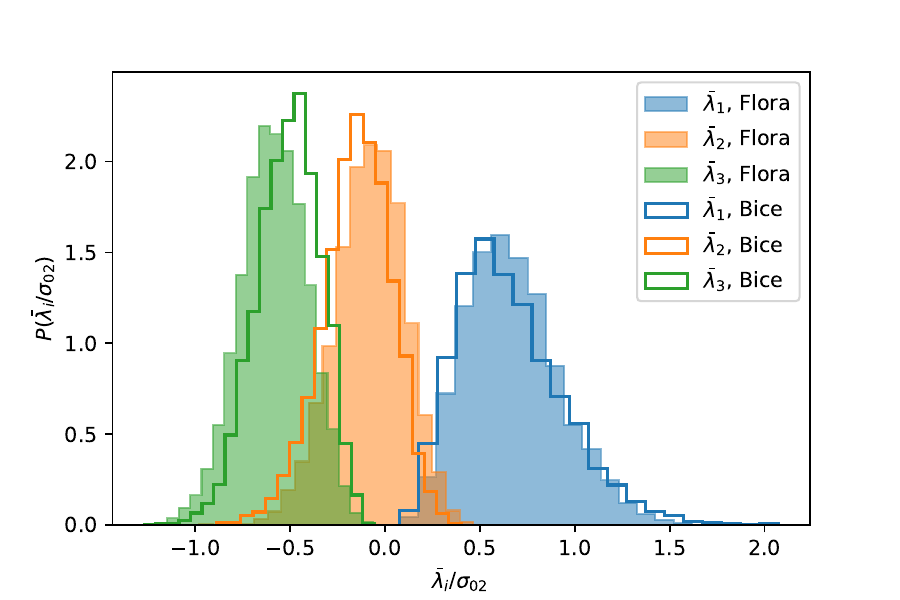}
\caption{\emph{Left panel.} Distribution of the (normalized) eigenvalues $\lambda_i$ of the potential energy overdensity tensor.  Filled and open histograms show results for all sampled protohaloes having masses larger than $4\times10^{14} h^{-1} M_\odot$ (from Flora) and larger than $10^{11} h^{-1} M_\odot$ (from Bice), respectively. \emph{Right panel.}  Same, but now for the eigenvalues $\bar\lambda_i$ of the traceless part of the potential energy overdensity tensor.  Filled and open histograms are very similar, indicating that differences in the left panet are mainly driven by the mass-dependence of the trace.}
\label{fig:hist_eps}

\end{figure*}

We now test our \emph{ansatz} on the positivity of the eigenvalues of the energy overdensity tensor in the protohaloes of two simulations from the SBARBINE suite \citep{despali16}, namely Bice and Flora. Both simulations contain $1024^3$ dark matter particles in a periodic box of side $L_{\mathrm{box}}=125h^{-1}$Mpc and $L_{\mathrm{box}}=2h^{-1}$Gpc, with initial conditions generated at $z=99$. The corresponding particle masses are $1.55\times10^{8} h^{-1}M_\odot$ for Bice and $6.35\times10^{11} h^{-1}M_\odot$ for Flora. The simulations adopt a Planck13 background cosmology: $\Omega_m = 0.307$, $\Omega_\Lambda = 0.693$, $\sigma_8 = 0.829$ and $h = 0.677$.

Haloes are identified at $z=0$ using a Spherical Overdensity halo finder with a density threshold of $319$ times the background density, corresponding to the virial overdensity. Our Flora sample contains 5378 haloes, distributed as follows: all the 1378 halos more massive than $10^{15}h^{-1}M_\odot$, 2000 randomly selected haloes with masses between $10^{14}$ and $10^{15}h^{-1}M_\odot$, and 2000 randomly selected haloes with masses between $4\times 10^{13}$ and $10^{14}h^{-1}M_\odot$. 
Our Bice sample contains 5000 randomly selected haloes with masses between $10^{11}$ and $4\times 10^{13}h^{-1}M_\odot$.
For each halo identified at $z=0$, we use `protohalo' to refer to the region occupied by that halo's particles in the initial conditions.

Since we are interested in initial quantities, we work with the first snapshot of the simulation. We measured each protohalo's center of mass position and velocity, $\qcm$ and $\bm{v}_\cm$, by averaging over all its particles. We then constructed estimators for its inertia and potential energy overdensity tensor, defined in equations~\eqref{eq:Iij_def} and~\eqref{eq:uij_def}:
\begin{align}
  \hat I_{ij} &\equiv \sum_{n=1}^{N_H} (\qq^{(n)}-\qcm)_i(\qq^{(n)} - \qcm)_j \\
  \hat u_{ij} &\equiv -3\, \frac{\sum_{n=1}^{N_H} (\qq^{(n)}-\qcm)_i(\bm{v}^{(n)} - \bm{v}_\cm)_j/fDH}{
  \sum_{n=1}^{N_H} |\qq^{(n)}-\qcm|^2} \,,
\label{eq:esteps}
\end{align}
where $D$ is the $\Lambda$CDM density perturbation growth factor (at the redshift $z$ of the snapshot), $f=\dd\ln D/\dd\ln a$, and $n$ runs over the $N_H$ protohalo particles. 
Note that $I_{ij}$ is not divided by $N_H$.
To obtain accelerations from velocities we used the Zel'dovich approximation, in which $\nabla\phi \simeq \vv/\dot D= \vv/fDH$.

\begin{figure*}
  \includegraphics[width=\columnwidth]{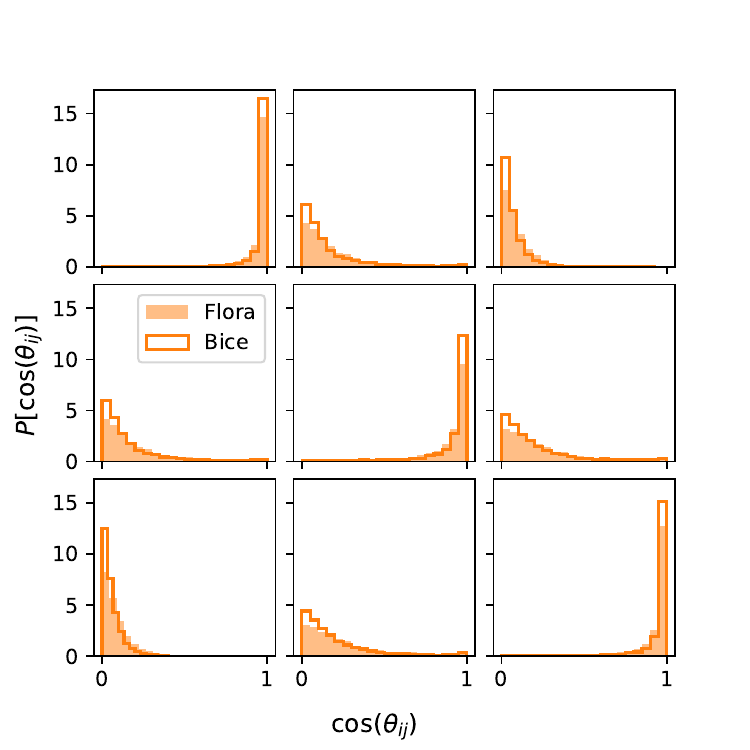}
  \includegraphics[width=\columnwidth]{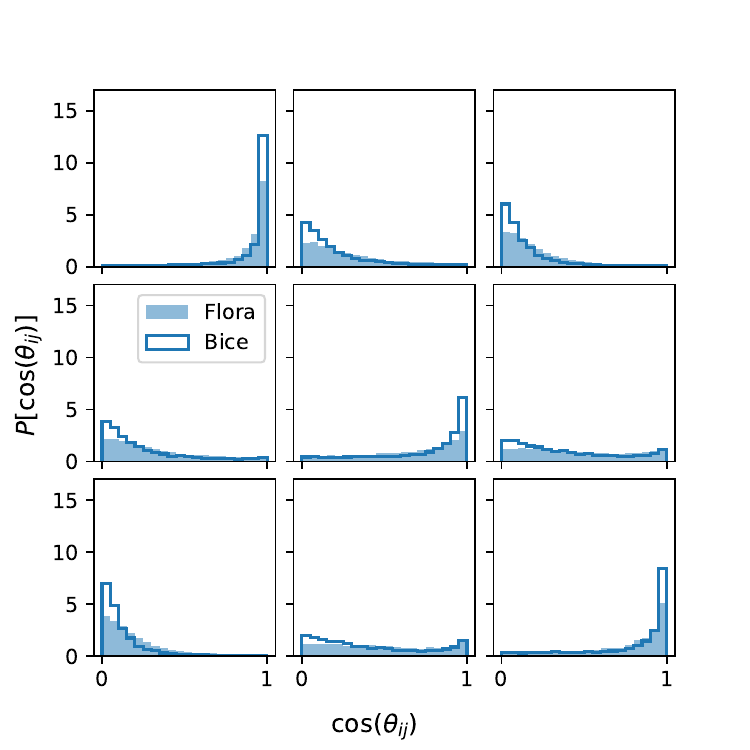}
\caption{Cosine of the angle $\theta_{ij}$ between the $i$-th eigenvector of the inertia tensor and the $j$-th one of the energy shear (left) or of the deformation tensor (right). The alignment with the energy shear is stronger.}
\label{fig:angles}
\end{figure*}

For each protohalo we also measure the mean deformation tensor, defined in equation \eqref{eq:deften} as the average over the protohalo particles of the second spatial gradient of the gravitational potential at each position. Since we need to take one more derivative, instead of getting the accelerations from the particle velocities (evaluated at irregular positions), we compute the gradient of the initial displacements that were imposed to create the initial conditions with N-GenIC \citep{Springel05}, which are computed on the grid. In practice, we select the grid points occupied by each protohalo, take their initial displacement field $\mathbf{\Psi}$, and estimate the mean deformation tensor \eqref{eq:deften} as:
\begin{equation}
  \hat{q}_{ij} \equiv \frac{1}{N_{G}}\sum_{k=1}^{N_{G}} \frac{\pd\Psi_j}{\pd q_i} \bigg|_k \, ,
\end{equation}
where $N_{G}$ is the number of grid cells contained within the Lagrangian protohalo volume. This means that we also include empty grid cells (i.e. those not occupied by halo particles) that are located inside the Lagrangian volume, because they too contribute to the total potential field that acts on the protohalo. This slightly refines the measurement one obtains if one uses the particle grid points only \citep[see][for more details]{dts13}.

The left panel of Fig. \ref{fig:eigenvalues} shows the eigenvalues $\lambda_1$, $\lambda_2$ and $\lambda_3$ of $\hat u_{ij}+\hat u_{ji}$ as a function of halo mass.  In the range of masses that we sampled, only one out of the 5378 Flora protohaloes had one slightly negative eigenvalue. For Bice protohaloes, which extend to much lower masses, this fraction rises to 0.5\% including only haloes mor massive than $10^{12} h^{-1} M_\odot$, and to nearly 5\% for all haloes in the sample.  Nevertheless, this is substantially smaller than the fractions shown in \cite{dts13} for density-based statistics -- i.e. the eigenvalues of $\hat{q}_{ij}$ -- which we also show in the right panel of Fig. \ref{fig:eigenvalues}. This fraction amounts to about 40 \% of all protohaloes, and nearly half of those between $10^{11} h^{-1} M_\odot$ and $10^{12} h^{-1} M_\odot$. Even at masses close to $10^{14} h^{-1} M_\odot$ one of the eigenvalues may still be negative.
This obvious difference, that the energy shear tensor is positive-definite even when the deformation tensor is not, is the main result of this paper.  

The left panel of Fig. \ref{fig:hist_eps} shows the distributions of the three eigenvalues of $\hat u_{ij}+\hat u_{ji}$, scaled by $\sigma_{02}$ defined in equation \eqref{eq:moments}, in the two simulations: filled histograms are for Flora protohaloes more massive than $4\times10^{14} h^{-1} M_\odot$, empty ones are for Bice protohaloes more massive than $10^{11} h^{-1} M_\odot$.  The less massive objects clearly have smaller $\lambda_i/\sigma_{02}$.  This is not so surprising, since we know that the trace, $\epsilon=\sum_i\lambda_i$, depends on mass \citep{epeaks}. More precisely, whereas the unnormalized $\epsilon$ increases at smaller masses, $\epsilon/\sigma_{02}$ tends to decrease, which explains why all the $\lambda_i/\sigma_{02}$ tend to be smaller. To remove this dependence, the right panel of Fig.~\ref{fig:hist_eps} shows -- for the same objects -- the distributions of the eigenvalues of the traceless part of the potential energy overdensity tensor, 
\begin{equation}
  \bar{\lambda}_i \equiv \lambda_i - \epsilon/3, 
\end{equation}
again normalized by $\sigma_{02}$ (same color coding). Now the histograms for the two simulations (filled and empty) are very similar, showing that these distributions are nearly mass independent.

\begin{figure}
  \includegraphics[width=\columnwidth]{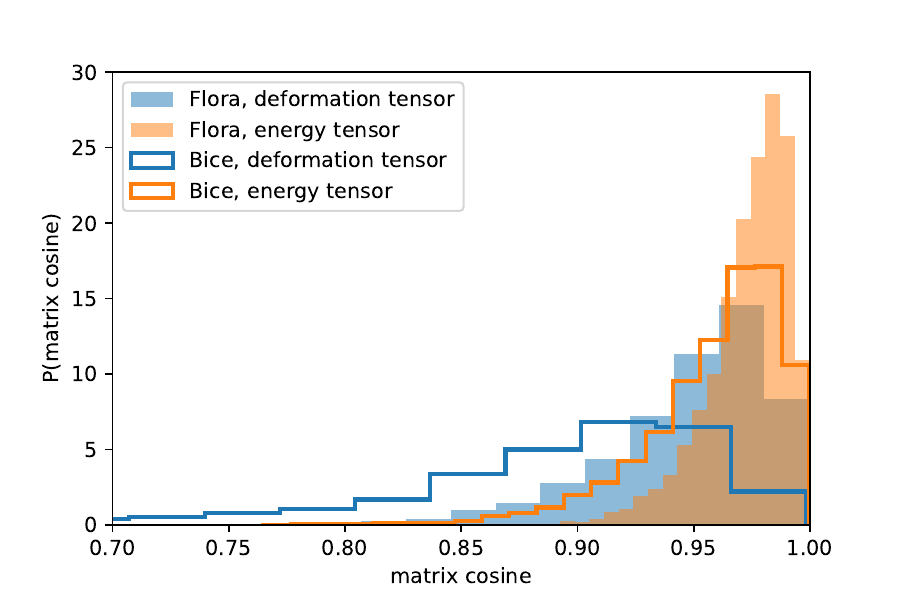}
\caption{Distribution of the matrix cosine (equation \ref{eq:matcos}) of the inertia tensor $I_{ij}$ with the energy shear tensor $u_{ij}$ (orange) or the deformation tensor $q_{ij}$ (blue). Filled histograms are for Flora (massive halos) and empty ones for Bice (smaller masses).}
\label{fig:cosine}
\end{figure}

\begin{figure}
    \centering
    \includegraphics[width=\columnwidth]{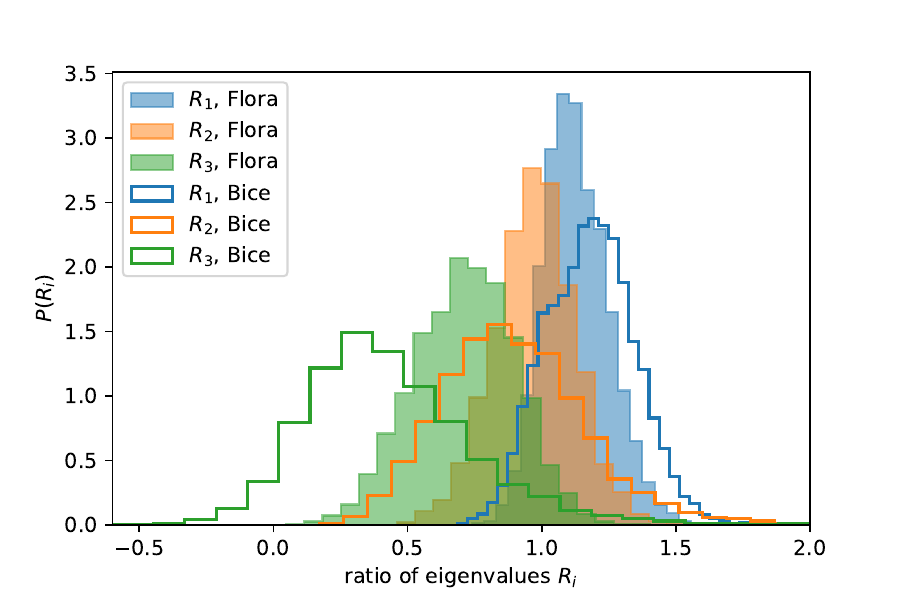}
    \caption{Ratios of the $i$-th eigenvalue $\lambda_i$ of the energy shear $u_{ij}$ to the $i$-th eigenvalue $a_i^2$ of the inertia tensor $I_{ij}$, both normalized to their traces: $R_i\equiv(\lambda_i/\epsilon)/(a_i^2/I)$. Equation \eqref{eq:align} would predict this ratio to be 1, whereas on average $R_1>1$ and $R_3<1$, meaning that protohaloes are slightly less elongated than this simplest prediction.}
    \label{fig:eigen_ratios}
\end{figure}

A detailed analysis of the mutual alignment of the three eigenvectors of the inertia tensor with those of the energy shear is shown in the panels on the left of Fig. \ref{fig:angles}. This alignment is quite strong, and stronger than the one with the deformation tensor, shown for comparison in the panels on the right. Interestingly, in both cases the alignment is better in Bice (i.e. for smaller halo masses) than in Flora (larger masses).

To get a more concise description of the overlap of the inertia tensor with the energy shear, for each protohalo we computed the matrix cosine
\begin{equation}
  \cos(\hat u,\hat I) \equiv 
  \hat u_{ij}\hat I_{ji}/\sqrt{\hat u_{kl}\hat u_{lk} \hat I_{mn}\hat I_{mn}}\,,
  \label{eq:matcos}
\end{equation}
which quantifies with a single number the alignment between the eigenvectors of the two matrices and the similarity of their eigenvalues. We repeated the same measurement for the deformation tensor. A value of the cosine closer to 1 indicates stronger alignment, and is what equation \eqref{eq:align} would predict exactly. The orange histograms in Fig.~\ref{fig:cosine} show that most measured cosines are greater than 0.9, although the alignment is slightly better for massive halos (filled histogram).  For comparison, the blue histograms show the result of replacing $\hat u$ with the deformation tensor $\hat q$:  clearly, the inertia tensor is better aligned with the energy shear than with the deformation tensor.

The direct comparison of the eigenvalues $a_i^2$ of the inertia tensor and $\lambda_i$ of the energy shear, each normalized to their sum $I$ and $\epsilon$, is shown in Fig. \ref{fig:eigen_ratios}, which displays their ratios $R_i\equiv(\lambda_i/\epsilon)(I/a_i^2)$. Equation \eqref{eq:align} would predict that the two tensors are proportional to each other, and therefore each ratio should be equal to 1. This prediction is clearly not verified. Fig. \eqref{fig:eigen_ratios} suggests instead that protohaloes are slightly less elongated than they would have been if their inertia tensor and energy shear were proportional.

\begin{figure}
  \includegraphics[width=\columnwidth]{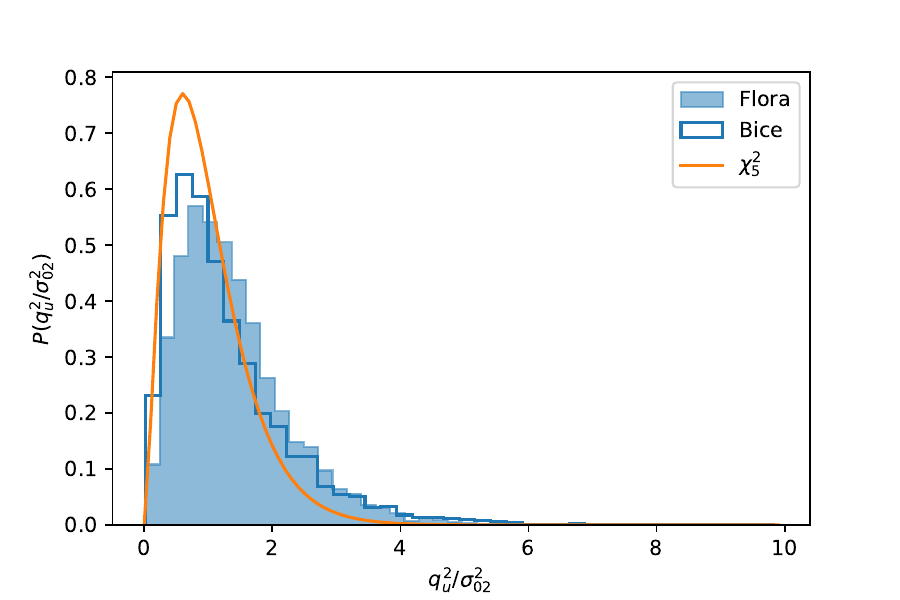}
\caption{Distribution of $q_u^2/\sigma_{02}^2$ for Flora (filled) and Bice (open) protohaloes; smooth curve shows the expected distribution for unconstrained positions (a $\chi^2$ distribution with 5 degrees of freedom).}
\label{fig:hist_u2}
\end{figure}

The histograms in Fig. \ref{fig:hist_u2} show the distribution measured in protohalos (in Flora and Bice respectively) of the shear strength $q_u^2/\sigma_{02}^2$ defined in equation \eqref{eq:qu2}. This distribution is broader than a $\chi^2_5$, shown for comparison by the smooth solid curve, which would be expected if the eigenvalue of the energy overdensity tensor were free to take up any sign. Nevertheless, the it is still rather mass independent.

\begin{figure}
    \centering
    \includegraphics[width=\columnwidth]{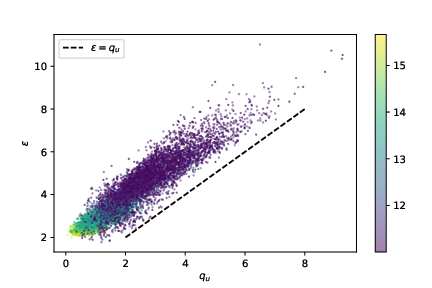}
    \caption{Correlation between the energy overdensity $\epsilon$ and the traceless energy shear $u$ of protohaloes. The color coding displays the (logarithm of the) halo mass. Dashed line shows the limiting value ($\epsilon=q_u$).}
    \label{fig:eps_vs_u}
\end{figure}

A scatter plot of $\epsilon$ vs $q_u$ (colored by protohalo mass) is displayed in Fig. \ref{fig:eps_vs_u}, and shows a strong degree of (approximately linear) correlation between the two variables.  This behaviour is also a consequence of the eigenvalues of $u_{ij}$ being all positive, as argued in Section \ref{sec:qu}.

\section{Comparison with previous work}
\label{sec:comparison}

Our analysis bears some qualitative similarity with the results of \cite{bm96}, but with two important differences. First, they considered an initially spherical region, in which case $AA^T$ in equation \eqref{eq:Iij} would be diagonal and reduce to $R^2\delta_{ij}$, where $R$ is the Lagrangian radius of the sphere. While this approximation can be useful and significantly simplifies the calculations, protohalos are not spherical (or even ellipsoidal), but have rather irregular shapes. 

Second, they focused on the deformation tensor $q_{ij}$,  rather than on $u_{ij}$. This is correct only under the assumption (which they make) that $V$ is spherical (or at least ellipsoidal) and homogeneous, in which case the two tensors are identical. 
Under this assumption, there is an exact analytical expression for the potential of the ellipsoid. Working with $u_{ij}$ is on the other hand completely general -- albeit based on a perturbative expansion -- and is therefore the correct choice for describing the collapse of the inertia tensor $I_{ij}$. The downside is that $u_{ij}$ has no closed analytical expression in terms of $I_{ij}$ (to be expected, since it also depends on the density profile, except for particular, highly symmetrical cases).

\cite{PorcianiTTTal2002}, and also later \cite{porciani11}, dropped the spherical protohalo assumption, but continued to assume that the deformation tensor provides the natural framework for understanding evolution.  In particular, they assumed that protohaloes are ellipsoidal with the longest axis being exactly aligned with the direction of maximum compression of the deformation tensor.  While this assumption works well in practice, our analysis here suggests that it has no fundamental physical justification.  Rather, its accuracy is more of statistical origin (and is not as good as the one with the energy shear). Furthermore, an exact alignment would not generate angular momentum via gravitational coupling, since the torque induced by the external gravitational field on the protohalo is zero. 

On the other hand, as shown by \cite{eshape23}, protohaloes are well approximated by equipotential surfaces that deviate from spherical symmetry in response to the local anisotropy of the infall potential. This gives an independent prescription for determining the initial protohalo shape, which then evolves under the action of $u_{ij}$. The symmetric part of $u_{ij}$ determines the collapse and deformation, and the antisymmetric one the torque \citep{eshape23}, which can then generate angular momentum as described by \cite{CadiouSmith2021}.

\cite{dts13} studied the eigenvalues of the deformation tensor, and noted that at lower masses (below $10^{13} h^{-1} M_\odot$) it is common for one of them to be negative. In hindsight, this is not surprising since in a generic setup the average of $\nabla_i\nabla_j\phi$ (that is, $q_{ij}$) describes the gradient of the center of mass acceleration (see discussion at the end of Section~\ref{sec:evolveI}), which is not directly implicated in the collapse dynamics. Rather, as we argue here, triaxial contraction follows from the positivity of the eigenvalues of the energy shear $u_{ij}$, which our measurements in Fig.~\ref{fig:eigenvalues} confirm.

Working with $u_{ij}$ has also practical advantages. For a $\Lambda$CDM power spectrum the variance of $\nabla_i\nabla_j\epsilon_R$ remains finite, whereas that of $\nabla_i\nabla_j\delta_R$ diverges. 
For $\epsilon_R$ it is thus possible to carry out the full  calculation of the peak number density from first principles, without having to resort to hand-waving changes of the smoothing filter like in a $\delta_R$-based approach \citep{epeaks}. However, one might also be interested in other statistics, like the number density of points with null determinant. These are the so-called critical events \citep{Hanami}, where an eigenvalue changes sign, because -- for instance -- a peak and a saddle point merge and disappear as the smoothing scale $R$ changes. These events are Lagrangian proxies for halo mergers \citep{critev}.
Computing their number density would require taking one extra derivative, but the variance of the third derivative diverges, this time also for $\epsilon_R$. On the other hand, if the eigenvalue changing sign is that of $u_{ij}$, the calculation remains well behaved. Defining critical events in terms of the energy tensor, rather than the Hessian, would thus not only come with a stronger connection with dynamics, but also better mathematical properties.

\section{Conclusions}\label{sec:final}
Our measurements strongly support the conclusion that protohalo patches can be identified with regions where the energy overdensity tensor is positive definite. This fact is not a simple statistical correlation, but a dynamical feature that follows directly from the evolution equations of the inertia tensor, which must collapse along three axes. On the other hand, the failure of the deformation tensor defined by the mass overdensity to predict triaxial collapse also has an explanation. While this tensor is directly related to the change of shape of a microscopic volume element, for a macroscopic body its physical interpretation is rather that of the gradient of the center of mass acceleration: therefore, not a quantity directly related to the collapse dynamics.

We have thus added another ingredient to the recipe for characterizing protohaloes in terms of energy-related quantities:  
\begin{itemize}
    \item the center of mass of a protohalo is much better identified by a local maximum of the energy overdensity field than of the matter overdensity \citep{epeaks}; 
    \item protohalo shapes and boundaries can be described by equipotential surfaces, which further maximize the enclosed energy with respect to spherical configurations \citep{eshape23};
    \item the energy overdensity tensor (whose eigenvalues must be all positive) does a much better job at predicting both the collapse of the three axes and their initial orientation than the deformation tensor (this work).
\end{itemize}

More specifically, we have shown that the principal axes of the inertia tensor of protohaloes are very well aligned with the eigenvectors of the energy tensor; the latter identify the main directions of compression of the protohalo region. The longest/shortest protohalo axis tends to align with the direction of maximum/minimum compression, corresponding to the largest/smallest eigenvalue of the energy tensor, for which infall is favoured/disfavoured. Moreover, no collapse should be possible if one eigenvalue is negative. Our measurements strongly support this prediction: less than 5\% of protohaloes down to $10^{11} h^{-1}M_\odot$ have one negative eigenvalue (and even then, only slightly negative); none has two negative values.

This is important for several reasons. First of all, it gives an effective and unambiguous description of the protohalo environment: protohaloes can be identified with regions with positive definite energy overdensity tensor, as this is what triggers triaxial collapse. Secondly, while the trace of the tensor (the energy overdensity $\epsilon$) sets the time of collapse and therefore determines the current halo mass, its traceless part (the traceless energy shear) provides a set of variables that can be used to describe assembly bias. It can incorporate effects of the anisotropy of the environment, \cite[as done e.g. by][using the tidal deformation tensor, the analogous quantity for the density]{st02,bweb18}, and would appear naturally in symmetry-based bias expansions \citep{scs13, genLagbias, djs18}. 

In fact, in our formulation, the energy shear is expected to source the first correction to the collapse time of $R_I$, and hence to appear in any analytical expression of a critical threshold for $\epsilon$.  Fig. \ref{fig:hist_u2} shows that, simply by working with $q_u^2/\sigma_{02}^2$, the mass-dependence can be nearly scaled-out; this should vastly simplify the development of models of the effects of anisotropies on halo collapse.
Furthermore, our analysis also shows (see Fig. \ref{fig:eps_vs_u}) that the positivity of the eigenvalues introduces a strong correlation between $\epsilon$ and the traceless shear (which would be independent if $u_{ij}$ were a generic Gaussian matrix, or equivalently, at random locations in the Gaussian field). This can help to explain the scatter and the trend with mass of the measured values of $\epsilon$ in protohaloes \citep{epeaks}, thus enabling the construction of a physically motivated model of the collapse threshold in the energy-based approach.  This would provide a natural way for the approach to include assembly bias effects.

There are also a number of practical implications for analytical models of halo aboundance and of the cosmic web. For instance, halos are typically identified with peaks in the (matter or) energy overdensity field.  To distinguish between peaks and other stationary points, one requires that the Hessian of the field be positive definite \citep{bbks86}.  However, one can replace this constraint on the Hessian with the same requirement on the energy overdensity tensor. Since the two tensors are correlated, the Hessian would then acquire a mean value proportional to $-\gamma u_{ij}$, where $\gamma$ denotes the normalized correlation coefficient between $-\nabla^2\epsilon$ and $\epsilon$. Thus, it is likely that if $u_{ij}$ is positive definite, then so is the negative Hessian, and the point is a peak. 

For a more quantitative grasp of the effect, note that requiring positive definiteness shifts the mean of $\epsilon$ from zero to $\langle\epsilon|+\rangle \approx 1.65\sigma_{02}$ \cite[see, e.g.,][]{ls98}.  The trace of the Hessian $x$ is commonly used to denote the peak `curvature'.  If $\gamma$ denotes the normalized correlation coefficient between $x$ and $\epsilon$, requiring positive definiteness of $\epsilon$ shifts the mean of $x$ to $\langle x|+\rangle/\sigma_{22}\approx \gamma\,\langle\epsilon|+\rangle/\sigma_{02}\approx 1.65\gamma$; i.e., the trace of the Hessian is likely to be positive, so the Hessian itself is likely to be positive definite, without explicitly requiring it to be so. 

Besides the Hessian, which requires two derivatives of the field, some models of the cosmic web make use of third derivatives to identify the disappearance of peaks (or other stationary points) via mergers \citep{Hanami, critev}.  In a $\Lambda$CDM scenario, the energy-based approach is attractive because the statistics of its second derivatives depend on integrals that do not diverge.  However, the statistics of its third derivatives still diverge, so, to estimate abundances of mergers will require using a different (essentially ad hoc) filter.  Our results suggest that more principled progress can be made by replacing the Hessian of the energy overdensity field with the energy tensor. This would enable estimation of the relevant statistics (which now have the same degree of convergence in $k$-space as a gradient) without mathematical ambiguities.   I.e., a constraint based on actual collapse dynamics may be better suited for this task than a merely topological one, and, like for energy peaks, considering a better-motivated physical quantity also solves the mathematical issues in the calculation of the statistics.

The idea of replacing the role of the Hessian with the energy overdensity tensor can be naturally extended to other constituents of the cosmic web \citep{bkp96}.  The signature of the eigenvalues of the energy tensor could be used as the fundamental marker that distinguishes between different cosmic web environments: the four possible different signatures of the potential energy tensor should be used to characterize nodes, filaments, walls and voids. In all these cases the underlying physical interpretation of the energy tensor, based on predicting the collapse vs expansion of the various axes, would be the same. It would be interesting to investigate how well this approach could reproduce properties of the cosmic web that are usually studies with density-based methods, such as its skeleton \citep{skeleton2008} or connectivity \citep{Codis_connectivity2018}.

A classification of the constituents of the cosmic web based on the dynamical role of the energy shear would connect with the well-established literature on using the signature of the deformation tensor (i.e., the Hessian of the potential rather than of the density) to characterize the cosmic web \cite[e.g.][but also, e.g., \citealt{sams06}]{hpcd07,Feldbrugge_tidal2023}.  Our work suggests that using the potential energy tensor instead should be more robust.  Perhaps more importantly, as Figure~\ref{fig:eigenvalues} shows, the signatures of the potential energy and deformation tensors can be different, so the actual cosmic web classifications will differ. 
Another obvious connection would be with the various works that have tried to predict the emergence of the cosmic web from the formation of caustics in the velocity field \citep{arnold_shandarin_zeldovich1982,Hidding2013, origami2014,Feldbrugge_caustics2017}. Unlike these works, which investigate the microscopic properties of the velocity field, the approach based on the energy shear attempts to predict the evolution of macroscopic volumes. It is therefore more suitable to make analytical estimates of the statistics of objects.

Finally, as a practical matter, the measurement of $u_{ij}$ in simulations is very easy, since it only depends on positions and velocities of protohalo particles. In contrast, $q_{ij}$ requires computing the gradient of the displacement, which needs to be done at grid points and thus requires extra work.


\section*{Acknowledgements}

Thanks to Corentin Cadiou, Christophe Pichon and Dmitri Pogosyan for helpful discussions, and to the organisers and participants of the 2023 KITP Cosmic Web workshop. Our visit to KITP was supported by the National Science Foundation under PHY-1748958.
GD acknowledges the funding by the European Union - NextGenerationEU, in the framework of the HPC project – “National Centre for HPC, Big Data and Quantum Computing” (PNRR - M4C2 - I1.4 - CN00000013 – CUP J33C22001170001).
\section*{Data Availability}

The simulation data and post-processed quantities used in this work can be shared on reasonable request to the authors.



\bibliographystyle{mnras}
\bibliography{mybib}







\bsp	
\label{lastpage}
\end{document}